\documentclass{epl}
\usepackage{graphicx}

\title{Dynamics of a Rigid Rod in a Glassy Medium}
\author{Angel J. Moreno\inst{1,2} and  Walter Kob\inst{1}}
\institute{
\inst{1}Laboratoire des Verres. Universit\'{e} de Montpellier II. Place E. Bataillon.
CC 069. F-34095 Montpellier, France.\\
\inst{2}Dipartimento di Fisica and INFM
Udr and Center for Statistical Mechanics and Complexity. Universit\`{a}
di Roma ``La Sapienza''. Piazzale Aldo Moro 2. I-00185 Roma, Italy.
}
\pacs{66.30.-h}{Diffusion in solids}
\pacs{05.40.-a}{Fluctuation phenomena, random processes, noise, and Brownian motion}
\pacs{64.70 Pf}{Glass transitions}

\begin{document}

\maketitle
\vspace{-8 mm}
\begin{abstract}
We present simulations of the motion of a single rigid rod in a disordered
static 2d-array of disk-like obstacles.  The rotational, $D_{\rm R}$,
and center-of-mass translational, $D_{\rm CM}$, diffusion constants are
calculated for a wide range of rod length $L$ and density of obstacles
$\rho$. It is found that $D_{\rm CM}$ follows the behavior predicted by
kinetic theory for a hard disk with an effective radius $R(L)$.  A dynamic crossover
is observed in $D_{\rm R}$ for $L$ comparable to the typical distance
between neighboring obstacles $d_{\rm nn}$.  Using arguments from kinetic
theory and reptation, we rationalize the scaling laws, dynamic exponents,
and prefactors observed for $D_{\rm R}$.  In analogy with the enhanced
translational diffusion observed in deeply supercooled liquids, the
Stokes-Einstein-Debye relation is violated for $L > 0.6d_{\rm nn}$.
\end{abstract}
\vspace{-4 mm}
Lorentz gases are systems in which a {\it single} classical particle
moves through a disordered array of static obstacles. They are used
as simple models for the dynamics of a light particle in a disordered
environment which presents a much slower dynamics. With increasing density
of the obstacles, dynamic correlations and memory effects start to become
important for the motion of the diffusing particle, and the system shows
the typical features of the dynamics of supercooled liquids, such as
a transition to a non-ergodic phase of zero diffusivity \cite{gotze,bruin}.
The theoretical description of the dynamics of the Lorentz gas
is highly non-trivial since, e.g., the diffusion constant
or the correlation functions are non-analytical functions of the density
of obstacles \cite{gotze,bruin}.

Diffusing particles and obstacles are generally modeled as disks or
spheres in two and three dimensions, respectively. In this Letter we
use molecular dynamics simulations to investigate, at low and moderate
densities, a generalization of the Lorentz gas, namely a model in which
the diffusing particle is a rigid rod. In contrast to the usual models,
this system has an orientational degree of freedom, which for sufficiently
long rods is expected to be strongly slowed down by steric hindrance. 
Furthermore this system will allow to gain insight
into the relaxation dynamics of a nonspherical probe molecule immersed
in a simple liquid, an experimental technique that is used, e.g.,
in photobleaching experiments to study the dynamics of glass-forming
liquids~\cite{cicerone}, or the investigation of the
coupling between the rotational and translational degrees of freedom in
such liquids~\cite{andreozzitaschin}.

In the following we will consider for the array of obstacles a homogeneous
``glassy" configuration instead of the more usual random structure. In
this way we avoid the effects induced by the very different length scales
present in a completely random medium \cite{moreno}, and which complicate the
physical interpretation of the observed dynamic features.  For simplicity,
simulations have been done in two dimensions, reducing the degrees of
freedom to a rotational and two translational ones. However, we think that
the physical picture proposed here for the interpretation
of the the obtained results is also valid in a three dimensional system.

The rigid rod, of mass $M$, was modeled as $N$ aligned beads of equal
mass $m=M/N$, with a bond length $2\sigma$. The rod length is thus
given by $L=(2N-1)\sigma$. In order to obtain the glassy host medium,
we equilibrated at a density $\rho = 0.77\sigma^{-2}$ and at temperature
$T=\epsilon/k_{\rm B}$ a two-dimensional array of soft disks interacting
via a potential $V(r)=\epsilon(\sigma/r)^{12}$. This procedure produced an
homogeneous liquid-like configuration. The latter was then permanently
frozen and was expanded to obtain the desired density of
obstacles, defined as $\rho = n_{\rm obs}/l_{\rm box}^{2}$, with
$n_{\rm obs}$ the number of obstacles and $l_{\rm box}$ the length
of the square simulation box used for periodic boundary conditions.
The same soft-disk potential $V(r)$ was used for the interaction between
the beads and the obstacles.  For computational efficiency, $V(r)$
was truncated and shifted at a cutoff distance of 2.5$\sigma$. In
the following, space and time will be measured in the reduced units
$\sigma$ and $(\sigma^{2}m/\epsilon)^{1/2}$, respectively.
The rod was equilibrated at $T = \epsilon/k_{\rm B}$. After the
equilibration, a production run was done at constant energy. 
For statistical average, runs were carried out for
typically 600-1000 different realizations
of the ensemble single rod-obstacles. Runs covered $10^{6}$
time units, corresponding to $(1-5)\times 10^{8}$ time steps, depending on
the step size used for the different rod lengths and densities.  Run times
were significantly longer than the relaxation times of the system.

Fig. 1 shows typical trajectories of the rod center-of-mass (CM) for
$\rho = 10^{-2}$ and different rod lengths. A transition from a typical
random walk for short rods to a polygonal-like structure for long ones
is observed. The latter consists of nearly longitudinal motions inside
corridors delimited by the neighboring obstacles, connected by vertices
corresponding to regions where the rod changes from a corridor to another
one. The very different nature of these trajectories suggest a change
in the transport mechanism when increasing the rod length at constant
$\rho$. Thus, while for short rods simple rotational and translational
Brownian dynamics is expected, for long ones rotations and transversal
translations are strongly hindered, and one has reptation-like motion
inside tubes formed by neighboring obstacles and which gives rise to
the nearly straight long paths observed in the trajectories.

In order to understand this change in the dynamics we will make use of kinetic
theory and reptation arguments. Consider first the translational
dynamics. We describe the rod by an ``effective hard rod'' formed by a
rectangle of width $2\sigma$ and length $L-\sigma$ and two semi-circles
of radius $\sigma$ attached at both ends, leading to an effective rod
length $L+\sigma$. This allows us to set the size of the obstacles
to zero, i.e. to treat them as point particles. If $\psi$ is the angle
formed by the rod axis and the CM velocity, it is easy to see that
the projection of the effective rod in the transversal direction is
$2\sigma+(L-\sigma)\sin\psi$. Integration over all the orientations yields
an average projection $2R$, with $R=(L+(\pi-1)\sigma)/\pi$. The isotropic
nature of the Brownian regime at low densities suggests 
to substitute the real system by an equivalent Lorentz gas of hard disks
having this latter radius. Kinetic theory and mode coupling calculations
for hard disks of radius $R$ and mean velocity $\langle v\rangle$ in
a 2d-medium with density of point obstacles $\rho$ lead, in first order,
to the following result, valid at low and moderate densities \cite{gotze}:

\begin{equation}
D_{\rm CM}/D_{\rm CM}^{0} = 1+(32/9\pi)\rho^{\ast}\ln\rho^{\ast}\quad ,
\end{equation}
\noindent
with the reduced density $\rho^{\ast}=\rho R^{2}$, and $D_{\rm CM}^{0}
= 3\langle v\rangle/16\rho R$ is the low-density limit, corresponding to
Brownian motion -i.e., no correlation between consecutive collisions.
Since for a Maxwell-Boltzmann distribution the mean CM velocity of
the rod is given by $\langle v\rangle = (\pi k_{\rm B}T/2M)^{1/2}$,
the corresponding low-density limit for the CM translational diffusion
constant of the rod, within the picture of equivalent hard disks of
radius $R=(L+(\pi-1)\sigma)/\pi$, is:

\begin{equation}
D_{\rm CM}^{0} =
\frac{3}{16}\left(\frac{\pi^{3}k_{\rm B}T}{2M}\right)^{1/2}\frac{1}{\rho(L+(\pi-1)\sigma)}
\end{equation}
\noindent
and $D_{\rm CM}$ is obtained from Eqs. (1) and (2) by taking
$\rho^{\ast} = \rho(L+(\pi-1)\sigma)^{2}/\pi^{2}$.

Points in Fig. 2a correspond to the ratio $D_{\rm CM}/D_{\rm CM}^{0}$
obtained from the simulations, where $D_{\rm CM}$ is calculated
as the long-time limit of $\langle(\Delta r)^{2}\rangle/4t$, with
$\langle(\Delta r)^{2}\rangle$ the mean squared CM displacement.
Brackets denote the ensemble average. As can be seen in Fig. 2b, within the
$\rho-$ and $L-$range investigated, the $D_{\rm CM}$ investigated cover
about 4 orders of magnitude.
Also included in Fig. 2a, is the theoretical prediction given by Eqs. (1) and (2).
A good agreement with the simulation data is obtained for reduced densities $\rho^{\ast}$
smaller than $\xi_2 \approx 0.2$, thus showing that in that range, the picture
of the equivalent hard disk is correct even quantitatively and in particular,
the logarithmic corrections to $D_{\rm CM}^{0}$ in Eq. (1) give account
for the observed decrease of $D_{\rm CM}/D_{\rm CM}^{0}$
at intermediate values of $\rho^{\ast}$. Since the typical distance between neighboring
obstacles, $d_{\rm nn}$, is given by $d_{\rm nn}=\rho^{-1/2}$, $\rho^{\ast}=\xi_{2}$
corresponds to a rod length $L \approx 1.4d_{\rm nn}$, and to a diameter
for the equivalent hard disk of $2R \approx 0.9d_{\rm nn}$. For hard disks of this size,
one has very strong correlation effects and the first order expansion in Eq. (1) is not
valid any more \cite{gotze}. Indeed the continuation of Eq. (1) to larger values of
$\rho^{\ast}$ leads to an unphysical increase of the ratio $D_{\rm CM}/D_{\rm CM}^{0}$ (see 
the theoretical curve in Fig. 2a). In contrast to this, the increase of this ratio 
observed in the simulation
results {\it does} have a physical origin. (Note that this increase is found 
only in the ratio and that $D_{\rm CM}$ decreases with increasing $\rho^{\ast}$, see Fig.~2b.)
However, it is not possible to rationalize this increase within 
the picture of equivalent hard disks of radius $R=(L+(\pi-1)\sigma)/\pi$.
Instead, reptation of the effective
rod allows the system to stay ergodic, which would not be the case for hard disks at 
large $\rho^{\ast}$, and hence to increase the ratio beyond 1.0.
We will give below evidence for reptation as the transport mechanism
for large $\rho^{\ast}$.

For the case of the rotational dynamics, a calculation from
first principles, that would require to take into account the
interactions of the obstacles with all the beads forming the rod,
is not straightforward. However, it is well-known that heuristic
approximations in the framework of elementary kinetic theory \cite{mcquarrie},
as the one we are going to discuss now, lead, {\it for very low
densities}, i.e., in the Brownian regime, to correct results
within a factor $O(1)$.  For a given ensemble of $n_{\rm 0}$ rods
with CM velocity $v$ and angular velocity $\omega$, we calculate
$n(\theta)$, the number of rods that rotate a given angle $\theta$
without colliding. A differential equation for this quantity can be
obtained by taking into account that the number of rods undergoing
a collision between $\theta$ and $\theta+d\theta$ is $n(\theta)$
multiplied by the probability of a collision in $d\theta$. By taking
advantage of the equivalent hard disk picture, such a probability can
be obtained as $\rho$ times the area swept by the equivalent hard disk,
$2Rdx = 2(L+(\pi-1)\sigma)dx/\pi$, where $dx$ and $d\theta$ are related
through $dx=vd\theta/\omega$. Thus we obtain the differential equation
$n(\theta+d\theta)-n(\theta) = dn = -2n(\theta)\rho Rvd\theta/\omega$,
and after integration, $n(\theta)=n_{0}\exp(-2\rho Rv\theta/\omega)$.
The mean angular free path between collisions is obtained as
$\langle\theta\rangle=(1/n_{0}){\rm{\int}}dvf(v){\rm{\int}}f(\omega)d\omega
{\rm{\int}}^{n_{0}}_{0}\theta dn$, where $f(v)$ and $f(\omega)$
are the Maxwell-Boltzmann distributions at temperature $T$ of
the CM and angular velocities respectively.  Integration leads to
$\langle\theta\rangle= \langle\omega\rangle\langle 1/v\rangle/(2\rho R)$.
For a Maxwell-Boltzmann distribution we have $\langle\omega\rangle
=(2k_{\rm B}T/\pi I)^{1/2}$, with $I$ the moment of inertia of the rod,
and $\langle 1/v\rangle=(\pi M/2k_{\rm B}T)^{1/2}$.  Therefore, from all
these results, and the expression from kinetic theory for the rotational
diffusion constant, $D_{\rm R}=\langle\theta\rangle\langle\omega\rangle$,
we obtain:

\begin{equation}
\frac{D_{\rm R}I}{M^{1/2}[L+(\pi-1)\sigma]} =
\left (\frac{k_{\rm B}T}{2\pi^{3}}\right )^{1/2}\frac{1}{\rho^{\ast}}
\quad .
\end{equation}
\noindent
$D_{\rm R}$ is determined from the simulation as the long time limit
of $\langle(\Delta\phi)^{2}\rangle/2t$, with $\langle(\Delta\phi)^{2}\rangle$
the mean squared angular displacement. 
In Fig.~3 we show its dependence on $\rho^{\ast}$
(left set of data points and left ordinate).
The thick solid line for $\rho^{\ast} < \xi_{1} \approx 0.04$
represents Eq.~(3) and shows a good agreement with the simulation
results. This value of $\rho^{\ast}=\xi_{1}$ corresponds to a rod
length $L \approx 0.6d_{\rm nn}$. As a further test of the validity
of this heuristic approach, we can calculate the low-density limit
$D_{\rm CM}^{0}$ following the same arguments for the equivalent
system of hard disks, leading to a decay rate for a displacement $r$,
$n(r)=n_{0}\exp(-2\rho Rr)$, a mean free path $l = (2\rho R)^{-1}$,
and a diffusion constant $D_{\rm CM} = l\langle v\rangle/2 = \langle
v\rangle/4\rho R$, i.e., a factor which is only $4/3$ larger than
the rigorous result $D_{\rm CM}^{0} = 3\langle v\rangle/16\rho R$.
In Fig.~3 a crossover to a different dynamic regime is observed around $\rho^{\ast}
= \xi_{1}$.  As can be seen in Fig.~2, this value of $\rho^{\ast}$
corresponds to a decay of a 15\% from the low-density limit of $D_{\rm
CM}$, again consistent with the breakdown of the picture of Brownian dynamics.

For the regime $\rho^{\ast} > \xi_{1}$ we now follow reptation arguments
similar to those developed for solutions of rigid linear polymers
\cite{doi}.  Thus, we assume that at a given time, the hindrance effects
induced by the neighboring obstacles result in the confinement of the
rod in the center of a tube of width $S$ and length $L+\sigma$ equal
to that of the effective rod. We decompose the long-time rotational
diffusion as a sum of elementary processes $i$, each one having a 
persistence time $\tau_{i}$ inside a tube, where the rod sweeps
a typical angle $\Omega_{i}$.  The rotational diffusion constant is estimated
as $D_{\rm R} = \Omega^{2}/2\tau$, where the average values of $\Omega$ and $\tau$ are taken.
Due to the homogeneous nature of the host medium, only a small dispersion is
expected for $\Omega_{i}$ and $\tau_{i}$, and the latter equation should be 
a reasonable approximation.  The persistence time $\tau$ can be estimated
as the time that the rod takes to cover its effective length $L+\sigma$.
Once the rod has moved this distance, the old tube has been left and a new tube
is defined by the new neighboring obstacles.  Therefore, $\tau$ can be
obtained as $\tau = (L+\sigma)^{2}/4D^{0}_{\rm CM}$, with $D^{0}_{\rm CM}$
the low-density limit given by Eq. (2), since we are referring to the
translational dynamics of the rod {\it inside} the free space of the tube.
On the other hand, the angle swept inside the tube can be estimated as
$\Omega = 2S/(L+\sigma)$.  Due to the homogeneous configuration of the
host medium, the width $S$ can be estimated as $S = d_{\rm nn}-2\sigma =
\rho^{-1/2}-2\sigma$, where the term $2\sigma$ is an excluded area effect
due to the finite width of the effective rod. From all these results,
$D_{\rm R} = \Omega^{2}/2\tau$ follows the scaling law:

\begin{equation}
\frac{D_{\rm R}\left(1+\frac{\sigma}{L}\right)^{4}
M^{1/2}L}{(1-2\rho^{1/2}\sigma)^{2}\left(1+\frac{(\pi-1)\sigma}{L}\right)^{3}} =
\left(\frac{9k_{\rm B}T}{8\pi^{5}}\right)^{1/2}\frac{1}{(\rho^{\ast})^{2}}
\quad .
\end{equation}
\noindent
A good agreement with the simulation data is obtained, as shown by the
thick solid line in Fig. 3 for $\rho^{\ast} > \xi_{1}\approx 0.04$.
Only a small, though systematic, deviation is observed at the largest
values of $\rho^{\ast}$, suggesting that in this range the model is
somewhat oversimplified.  Except for these largest values, the agreement
is particularly good for $\rho^{\ast} > \xi_{2}\approx 0.2$, i.e., for the
range where reptation was suggested as mechanism of ergodicity restoring
for the translational motion. For $\xi_{1} < \rho^{\ast} <\xi_{2}$,
corresponding to a rod length $0.6d_{\rm nn} < L < 1.4d_{\rm nn}$,
reptation also gives a reasonable description of the simulation data,
although a more detailed theory might be required to describe this
crossover regime.

It must  be stressed that there is no functional continuity
between Eqs. (3) and (4) and hence a more rigorous model should
be introduced to obtain a unified description at both sides of the
crossover point $\rho^{\ast}=\xi_{1}$. Nevertheless, we find that
the curves for $D_{\rm R}$ predicted by (3) and (4) actually cross
at a value $\rho^{\ast} = \xi_{\rm co}$ in agreement with the value
of $\xi_{1} \approx 0.04$ obtained from the simulations.  It is
straightforward to see that this common value is given by $\xi_{\rm
co}=\pi^{-2}\{1+(\pi-1)(\sigma/L)\}^{2}\{2(\sigma/L) + (2M/3\pi
I)^{1/2}(1+(\sigma/L))^{2}L\}^{-2}$.  This expression can be simplified by
taking into account that, except for the highest investigated densities,
the data in the reptation regime fulfill the condition $\sigma \ll L$
and therefore one can make the approximation $I=ML^{2}/12$. Using these
two approximations we obtain for $\xi_{\rm co}$ the result $\rho^{\ast}
= (8\pi)^{-1} \approx 0.04$, in agreement with the observed value of
$\xi_{1}$. From all these results, at this point it can be concluded that
Brownian motion and reptation are the transport mechanisms respectively
below and above $\rho^{\ast}=\xi_{1}$, the crossover taking place when
the rod length is comparable to the typical distance between neighbouring
obstacles. It is also noteworthy of remark that a consistent description
has been achieved without any fit parameter.

Some insight about the coupling between the rotational and translational
degrees of freedom can be obtained by investigating the validity of
the Stokes-Einstein-Debye (SED) relation \cite{egelstaff} $D_{\rm CM}\tau_{R} = c$, where
$c$ is a constant and $\tau_{R}$ is the rotational relaxation time. We
define this latter quantity as the time at which the correlation
function $\langle\cos\Delta\phi(t)\rangle$ has decayed to $1/e$.
In the limit of small $\rho^{\ast}$, where collisions are infrequent,
one expects that the rod performs several full rotations between
two consecutive collisions. In this case, the physical mechanism for
the angular decorrelation is not the collisions but the free rotation
of the rod between them. Therefore, in this limit $\tau_{\rm R}$ can be
calculated from the relation $\cos(\langle\omega\rangle\tau_{\rm R}) =
1/e$, leading to $\tau_{\rm R} = (\pi I/2k_{\rm B}T)^{1/2}\cos^{-1}(1/e)$.
A good agreement of this latter equation with the simulation data has
been obtained for $\rho^{\ast} < \xi_{1}$, confirming the validity of the
hypothesis (see Ref. [5]). Then, together with the low-density limit of
$D_{\rm CM}$ given by Eq. (2), we obtain the following scaling law for
the SED-relation:

\begin{equation}
D_{\rm CM}^{0}\tau_{R}\rho\left[L+(\pi-1)\sigma\right](M/I)^{1/2}
= (3\pi^{2}/32)\cos^{-1}(1/e)
\quad .
\end{equation}
\noindent
As shown in Fig. 4, the simulation data follow nicely this equation
for $\rho^{\ast} < \xi_{1}$, i.e., for $L<0.6d_{\rm nn}$.  Data for
$\xi_{1} < \rho^{\ast} <\xi_{2}$ are systematically below the constant
value $(3\pi^{2}/32)\cos^{-1}(1/e)$, showing that the 
SED-relation is no more fulfilled in this range of $\rho^{\ast}$ (the 
reason for this is the presence of the logarithmic correction given in 
Eq.~(1)). Finally,
a crossover to a power law $\sim (\rho^{\ast})^{2.4}$ is obtained at
$\rho^{\ast} =\xi_{2}$. (Note that presently we do not have an explanation for this power-law.) 
Thus, the SED-relation is clearly
violated for rods that are significantly longer than $d_{\rm nn}$, in
analogy with the phenomenon of enhanced translational diffusion observed
in supercooled liquids -with temperature as control \mbox{parameter-,} close
to the glass transition \cite{fujara}, where the value of $c$ increases
with decreasing temperature, and that is explained by the different
averaging of the rotational and translational degrees of freedom
\cite{stillinger}. This interpretation is clearly evidenced by the trajectory of
the center-of-mass for $L=59$ in Fig. 1 and the picture of reptation
at large values of $\rho^{\ast}$, where due to steric hindrance, the
rod is able to perform only small rotations while it moves along nearly
straight long paths. In this way, reptation leads to an enhancement of the
translational decorrelation in comparison with the rotational one.  

In summary, we have shown that arguments from kinetic theory and
reptation are able to give a very good acount for the translational
and rotational dynamics of the considered generalization of the Lorentz
gas. In particular the analysis show that in this system the violation
of the Stokes-Einstein-Debye relation is due to the breakdown of the 
isotropy of the rotational dynamics, a result that will be useful for the 
interpretation of the experimental findings of this violation.

\acknowledgments

We thank E. Frey for useful discussions.  A.J.M. acknowledges a postdoctoral
grant from the Basque Government. Part of this work was supported by
the EC Human Potential Program under contract
HPRN-CT-2002-00307, DYGLAGEMEM.
%
%\vspace*{-7 mm}
%
%
\begin{figure}
\begin{center}
\resizebox{0.38\columnwidth}{!}{\includegraphics{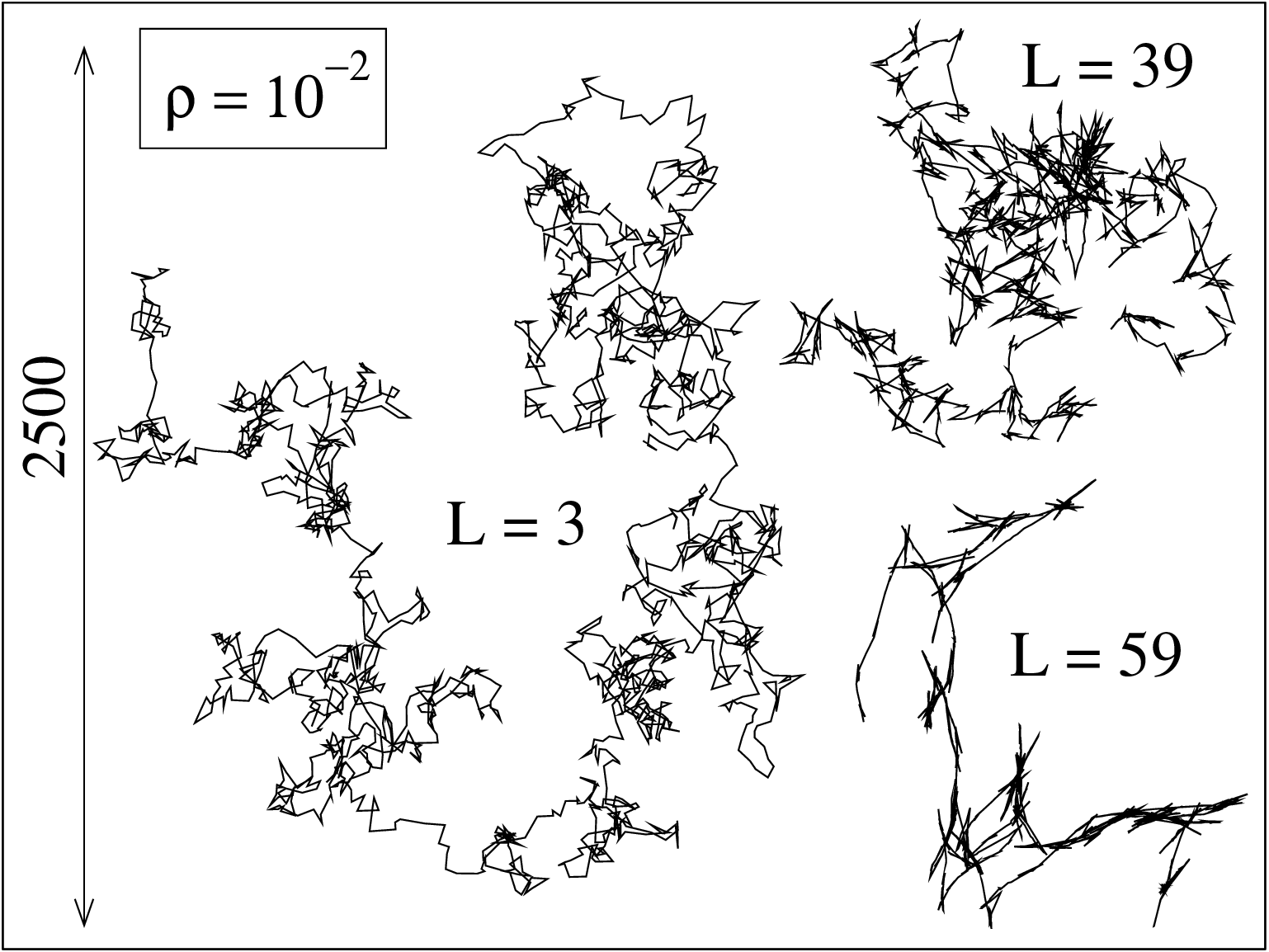}}
\end{center}
\caption{Typical trajectories of the rod CM at $\rho = 10^{-2}$, for $L=$
3, 39 and 59. The simulation time is $t_{\rm sim} = 10^{6}$, with a time
interval of 500 between consecutive plotted points.  The arrow indicates
the length scale. Note that for the sake of clarity the obstacles are
not shown.}
\label{fig1}
\end{figure}
%vspace{-0.3 cm}
%
%
\begin{figure}
\vspace{7 mm}
\begin{center}
\resizebox{0.66\columnwidth}{!}{\includegraphics{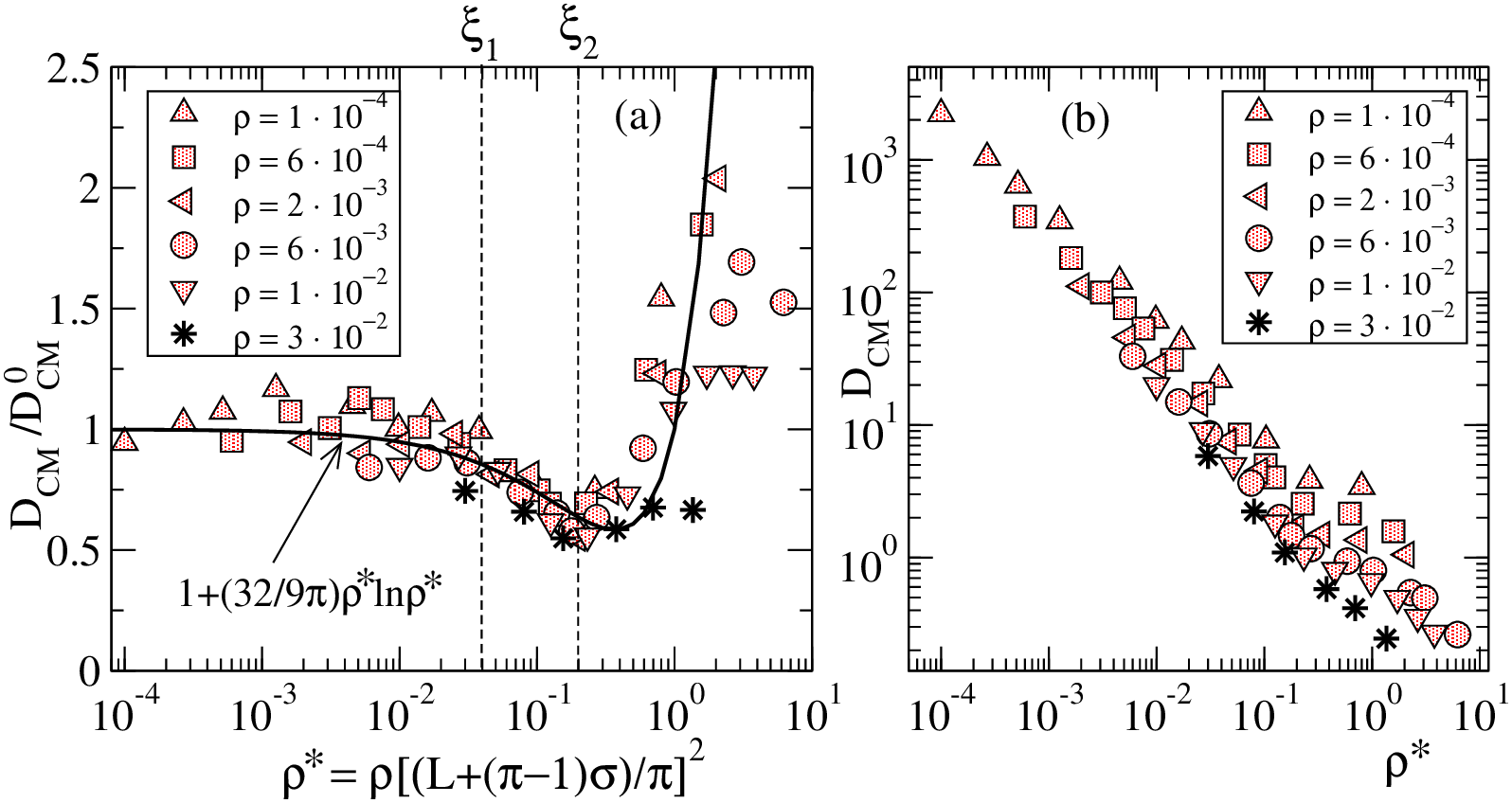}}
\end{center}
\caption{(a): Ratio between the CM translational diffusion constant
$D_{\rm CM}$ obtained from the simulations and the low-density limit 
$D_{\rm CM}^{0}$ for the equivalent hard disk (symbols).
The solid curve corresponds to the theoretical prediction of 
kinetic theory for the equivalent hard disk at low and moderate densities, see Eq.~(1).
(b): CM translational diffusion constant vs. the reduced density.}
\label{fig2new}
\end{figure}
%\vspace{-0.3 cm}
%
%
\begin{figure}
\begin{center}
\resizebox{0.43\columnwidth}{!}{\includegraphics{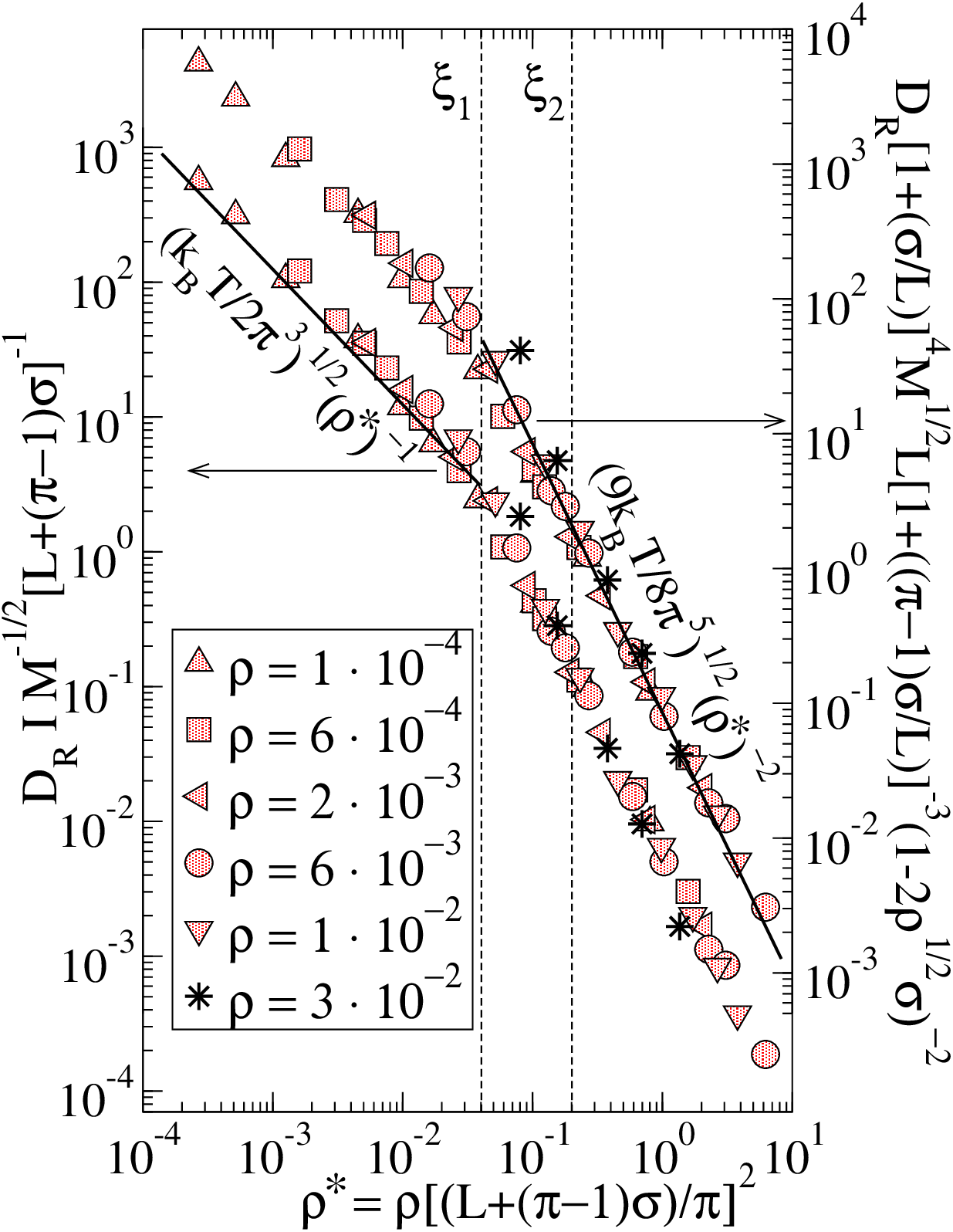}}
\end{center}
\caption{Scaling laws for $D_{\rm R}$. A double-$y$ representation
is shown for clarity.  Points are simulation data. Thick solid lines
correspond to the theoretical predictions for Brownian and reptation
dynamics (see text).}
\label{fig3}
\end{figure}
%\vspace{-0.3 cm}
%
\begin{figure}
\vspace{-3 mm}
\begin{center}
\resizebox{0.43\columnwidth}{!}{\includegraphics{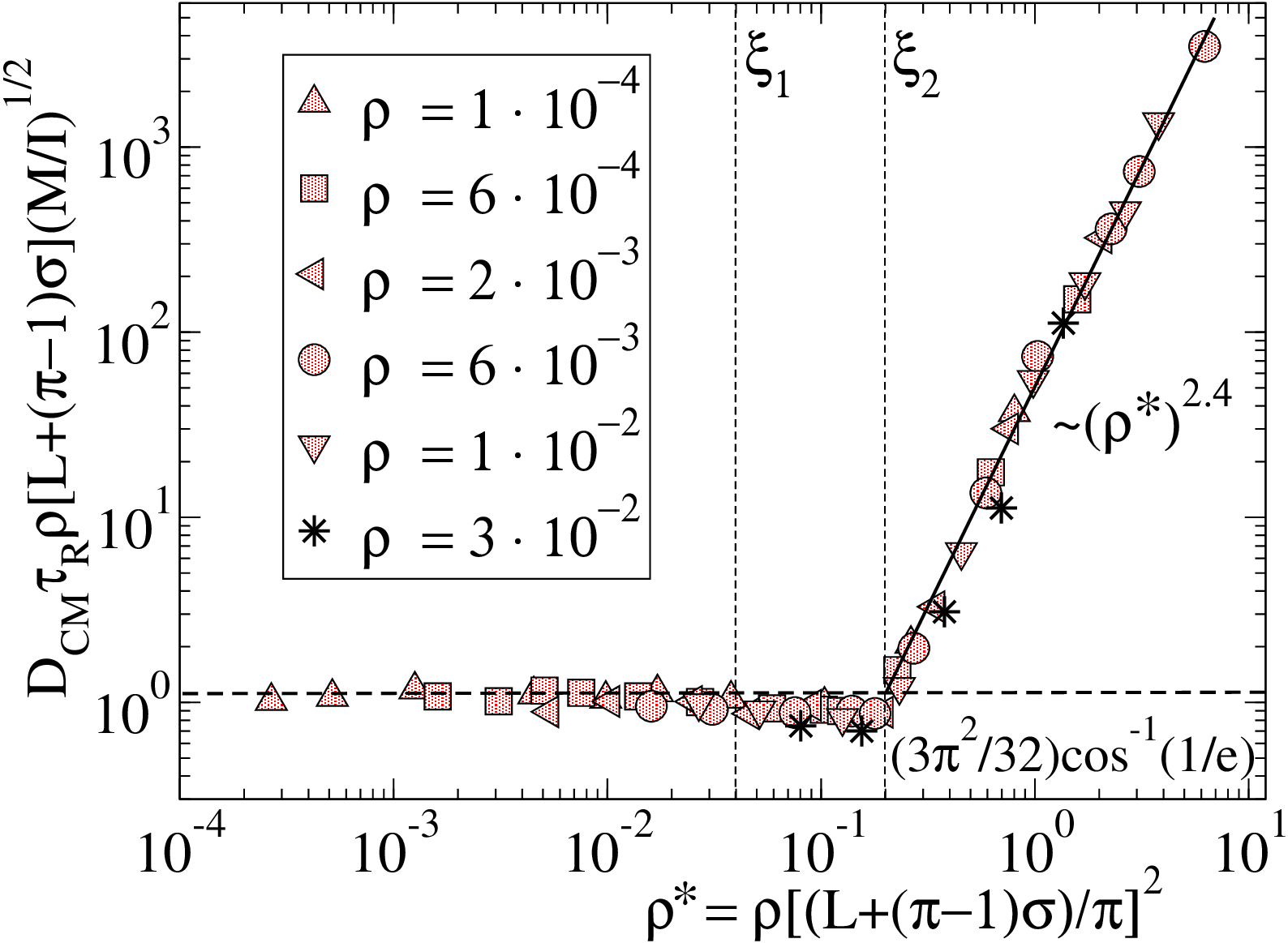}}
\end{center}
\caption{Scaling law for the product $D_{\rm CM}\tau_{\rm R}$ and violation of the
SED relation at large $\rho^{\ast}$.
The dashed line marks the theoretical constant value $(3\pi^{2}/32)\cos^{-1}(1/e)$ (see text).
The solid line in the large-$\rho^{\ast}$ regime is a fit to a power law.}
\label{fig4}
\end{figure}

\begin{thebibliography}{0}
%\vspace*{-16 mm}
%
\bibitem{gotze}
\Name{G\"{O}TZE W., LEUTHEUSSER E. and YIP S.}
\REVIEW{Phys. Rev. A}{23}{1981}{2634};
\REVIEW{ibid.}{24}{1981}{1008}.

\bibitem{bruin}
\Name{BRUIN C.}
\REVIEW{Phys. Rev. Lett.}{29}{1972}{1670};
\Name{MASTERS A. and KEYES T.}
\REVIEW{Phys. Rev. A}{26}{1982}{2129};
\Name{BINDER P. M. and FRENKEL D.}
\REVIEW{Phys. Rev. A}{42}{1990}{R2463}.

\bibitem{cicerone}
\Name{CICERONE M. T., BLACKBURN F. R. and EDIGER M. D.}
\REVIEW{J. Chem. Phys.}{102}{1995}{471};
\Name{CICERONE M. T., and EDIGER M. D.}
\REVIEW{J. Chem. Phys.}{103}{1995}{5684}.

\bibitem{andreozzitaschin}
\Name{ANDREOZZI L., DI SCHINO A., GIORDANO M. and LEPORINI D.}
\REVIEW{Europhys. Lett.}{38}{1997}{669};
\Name{TASCHIN A., TORRE R., RICCI M., SAMPOLI M., DREYFUS C. and PICK R. M.}
\REVIEW{Europhys. Lett.}{56}{2001}{407}.

\bibitem{moreno}
\Name{MORENO A. J. and KOB W.} 
\REVIEW{J. Chem. Phys.}{121}{2004}{380}.

\bibitem{mcquarrie} 
\Name{MCQUARRIE D. A.}
\Book{Statistical Physics}
\Publ{University Science Books, Sausalito, CA, USA}
\Year{2000}.

\bibitem{doi}
\Name{DOI M.}
\REVIEW{J. Physique}{36}{1975}{607};
\Name{DOI M. and EDWARDS S. F.}
\REVIEW{J. Chem. Soc., Faraday Trans.}{74}{1978}{560}.

\bibitem{egelstaff}
\Name{EGELSTAFF P. A.}
\Book{An Introduction to the Liquid State}
\Publ{Oxford University Press, Oxford, U.K.}
\Year{1994}.

\bibitem{fujara}
\Name{FUJARA F., GEIL B., SILLESCU H. and FLEISCHER G.}
\REVIEW{Z. Phys. B}{88}{1992}{195};
\Name{EDIGER M.D.}
\REVIEW{Ann. Rev. Phys. Chem.}{51}{2000}{99}.

\bibitem{stillinger} 
\Name{STILLINGER F. H. and HODGDON J. A.}
\REVIEW{Phys. Rev. E}{50}{1994}{2064}.
\end{thebibliography}
\end{document}